\documentclass[10pt,twocolumn,floatfix,superscriptaddress,amsmath,showkeys,aps]{revtex4-1}
\usepackage{bm}
\usepackage{xcolor}
\usepackage{hyperref}
\usepackage{makeidx}
\usepackage{amsmath}
\usepackage{graphicx}
\usepackage{dcolumn}
\usepackage{amsfonts}
\usepackage{amssymb}
\usepackage{color}
\usepackage{epstopdf}

\begin{document}

\title{Emergence of a non-Fowler-Nordheim-type behavior for a general planar tunneling barrier}

\author{Nei Lopes}
\email{nlsjunior12@gmail.com, neijr@cbpf.br (N. Lopes)}
\affiliation{Centro Brasileiro de Pesquisas F\'{\i}sicas, Rua Dr. Xavier Sigaud 150, Urca, 22290-180, Rio de Janeiro, Brazil}
\affiliation{Departamento de F\'{\i}sica, Universidade Estadual de Feira de Santana, Feira de Santana, 44036-900, Bahia, Brazil}
\author{A.V. Andrade-Neto}
\affiliation{Departamento de F\'{\i}sica, Universidade Estadual de Feira de Santana, Feira de Santana, 44036-900, Bahia, Brazil}

\date{\today}

\begin{abstract}

In this work we investigate a generalized tunneling barrier for planar emitters at \textit{zero-temperature}. We present the evidence of the emergence of a non-Fowler-Nordheim-type general behavior for the field emission current density in the case that the Fermi energy ($\mu$) is comparable with or smaller that the decay width ($d_F$). Therefore, for some non-metals or materials that have very small Fermi energy the standard Fowler-Nordheim-type theory may require a correction. In the opposite regime, i.e., for $\mu$ much larger that $d_F$, we confirm that the conventional theory is suitable for metals.\\
\textbf{keywords}: Field emission current density, Planar tunneling barrier, Fowler-Nordheim-type theory, small Fermi energy, non-Fowler-Nordheim-type behavior.\\

\end{abstract}

\maketitle

\section{Introduction}
\label{sec:Introduction}
Field emission (FE) is a process whereupon electrons are emitted from metallic materials due to the presence of a high electric field. This field narrows the bare potential energy (PE) barrier, such that the electrons present a non-negligible tunneling probability~\cite{Biswas}. This effect was first described by Fowler and Nordheim~\cite{Fowler} (FN). They considered a wave-mechanical tunneling through a triangular PE barrier but the image charge potential was not included. In $1956$, Murphy and Good~\cite{Murphy} (MG) introduced a more realistic PE barrier taking into account the image charge PE. They used the semi-classical (JWKB) approximation~\cite{Murphy, Cutler, Fursey, Neto2, Neto3} to obtain an expression for the barrier transmission coefficient. More recently, several works have been developed to include effects of tunneling, thermal emission and curvature for field emission in metals~\cite{Cutler, Jensen, Forbes4, Edgcombe, Fischer, Forbes6, Kyritsakis, Holgate}.

Since the 1960s, FE theoreticians have generally used MG theory~\cite{Murphy} to describe quantitatively the FE process from metals. However, this equation was derived within the context of planar emitters~\cite{Murphy, Edgcombe} and originally did not allow direct extraction of physical aspects associated with this phenomena since it was presented in terms of elliptic integrals, which implied that a numerical analysis~\cite{Dolan} was required. In that sense, many efforts have been devoted to improving the standard FN formula~\cite{Cutler, Jensen, Forbes4, Edgcombe, Fischer, Forbes6, Kyritsakis, Forbes1,Forbes2, Deane-Forbes} since the phenomena of field emission is an important ingredient for many technological applications, such as, lamp~\cite{Edgcombe}, X-ray sources~\cite{Edgcombe, Anantram}, field emission devices~\cite{Anantram}, and carbon nanotubes~\cite{Zestanakis, Filip, Mikhailovskij}. 

In this paper, we investigate the behavior of the FE current density at \textit{zero-temperature} within a generic planar tunneling barrier that does not depend on the exact form of the barrier. Without loss of generality, our treatment is focused on the materials with free-electron-like band-structure and no surface states. We use the semi-classical (JWKB) approximation~\cite{Murphy, Cutler, Fursey, Neto2, Neto3} to obtain the barrier transmission coefficient and recognizing that at \textit{zero-temperature} the only states occupied are those with negative or zero energy values relative to the Fermi level we expand the exponent of the transmission coefficient in a power series (Taylor-type) and truncate it in second order at the Fermi level. In order to obtain the FE current density formula we use the well-known method called integration via the normal energy distribution (NED). 

We study two distinct limits for the FE current density depending on the Fermi energy values and we present a general indication of the emergence of a non-FN-type behavior in planar emitters for the case that the Fermi energy ($\mu$) is comparable with or smaller that the decay width ($d_F$), that is, in the limit that the ratio $\alpha \equiv \mu/d_F<<1$, which implies that for some non-metals or materials that have very small Fermi energy, the standard FN-type theory may require a revision. In the opposite limit, i.e., for $\alpha \equiv \mu/d_F>>1$, we confirm that the conventional FN-theory is thus adequately valid for usual metals.

For completeness, it is worth to point out that this kind of non-FN-type behavior has been reported in different scenarios. For example, in highly curved, nanometer-scale surfaces~\cite{Holgate} as well as in multiwall carbon nanotubes~\cite{Zestanakis}. So, the context that gives rise to a non-FN-type behavior in general planar emitters clearly depends on the typical values for $\alpha$. For instance, for a usual metal with work function $4.5$ eV, $d_F$ typically has values in the range $100$ meV to $250$ meV, whereas for metals Fermi energies are typically around $5$ eV~\cite{Ashcroft-Mermin}, or more, which means that the conventional statement is appropriate for metals. On the other hand, for some non-metals, the Fermi energy is very much smaller. For such materials, or even for different materials that have small Fermi energy, if FN-type theory is used, as a first approximation, to describe FE from such materials, self-consistency may require that the emergence of a non-FN-type behavior presented here might be taken into account. Considerations of this general kind, i.e., related to the Fermi energy values, will also apply if the theory used is more general than conventional FE theory.

The paper is organized as follows: in Section~\ref{sec:FNformula} we present in details the main difficulty associated to extract some physical interpretations about the phenomena of field emission directly from the FN-type MG formula for planar emitters since it is expressed in terms of elliptic integral. In Section~\ref{sec:General_theory_for_planar_emitters} we describe a general approach for smooth planar emitters at \textit{zero-temperature}. We apply the semi-classical (JWKB) approximation to calculate the electronic transmission coefficient and expand its exponent about the Fermi level for derivate the current density formula to a planar tunneling barrier of any well-behaved shape. Then, we investigate two particular cases of interest for the current density depending on the ratio $\alpha=\mu/d_F$. Finally, in Section~\ref{sec:conclusions} we present our conclusions and summarize the main results.

\section{The FN-type MG formula for planar emitters}
\label{sec:FNformula}

In the standard FN-type \textit{zero-temperature} MG theory for planar emitters the FE current density ($J_{MG}$) is given by the well-known expression~\cite{Biswas, Cutler, Fursey, Edgcombe, Fischer, Haug, Forbes3, Forbes5},
\begin{equation}\label{eq1}
J_{MG}=\textit{t}_{MG}^{-2}a \phi^{-1}F^2\exp{\left[-\frac{\textit{v}_{MG} b \phi^{3/2}}{F}\right]}
\end{equation}
where $a$ and $b$ are universal constants known as first and second FN constants~\cite{Forbes5}, respectively. Moreover, the \textit{subscript} $MG$ here and elsewhere, indicates that the parameter refers to a Schottky-Nordheim (SN) barrier~\cite{Fischer} used in MG theory, $\phi$ is the local work function of the emitting surface, $F$ is the external electric field applied to narrow the bare PE barrier, which allows the electrons to tunnel out of the metal, and
\begin{equation}\label{eq2}
\textit{v}_{MG}=\left[\frac{1+\sqrt{1-f}}{2}\right]^{1/2}\left[E(k)-\left(1-\sqrt{1-f}\right)K(k)\right]
\end{equation}
whereupon $E(k)$ and $K(k)$ are complete elliptic integrals of the first and second kind, with
\begin{equation}\label{eq3}
k^2=\frac{2\sqrt{1-f}}{1+\sqrt{1-f}}
\end{equation}
where $k$ is the elliptic \textit{modulus} and
\begin{equation}\label{eq4}
f=c_s^{2}\phi^{-2}F=\frac{e^3}{4\pi \epsilon_{o}} \frac{F}{\phi^2}< 1
\end{equation}
where $c_s$ is the Schottky constant.

Finally, the special mathematical function \textit{t} can be expressed in terms of \textit{v} and $d\textit{v}/df$ as follows~\cite{Deane-Forbes},
\begin{equation}\label{eq5}
\textit{t}_{MG}(f)=\textit{v}(f)-\frac{4}{3}f\frac{d \textit{v}}{d f}
\end{equation}

According to all these definitions presented above, it is worth to emphasize that calculations of the current densities from Eq.~(\ref{eq1}) require an intrinsic numerical approach. Besides, note that it is very difficult to identify or even extract any physical interpretation about the FE current density direct from the FN-type MG formula, i.e., without a numerical treatment. Numerical calculations for typical values of external electric fields and work functions were performed in Ref.~\cite{Dolan}. Recent developments in relation to the standard FN theory are given in Refs.~\cite{Cutler, Jensen, Forbes4, Edgcombe, Fischer, Forbes6, Kyritsakis, Holgate, Forbes1, Forbes2, Deane-Forbes}.

\section{General theory for planar emitters}
\label{sec:General_theory_for_planar_emitters}

From now on, we present the evidence of the emergence of a non-FN-type behavior in the FE current density on the general theory for planar emitters in the small Fermi energies regime, i.e., when the values of Fermi energy is comparable with the decay width ($d_F$) of the \textit{zero-temperature} NED, aiming to characterize both the different regimes and some physical properties about the FE phenomena in the context of a planar tunneling barrier.

We will describe the generic result that can be applied to a generalized tunneling barrier of any well-behaved shape within the context of smooth planar emitters, i.e., which does not depend on the specific form of the barrier. Then, we will point out the scenario that gives rise to a non-FN-type behavior.

It is well-known that one can obtain the expression for the FE current density ($J$) by means of integrating the NED~\cite{Fowler_book}. Thus, $J$ can be written in the form~\cite{Fowler_book},
\begin{equation}\label{eq: current_density_exp}
J=\int j(E_n)dE_n= \int N_i(E_n)D(E_n)dE_n
\end{equation}
where $E_n$ is the normal energy measured relative to the base of the conduction band, $D(E_n)$ is the PE barrier transmission coefficient, and $N_i(E_n)$ is the incident NED. In addition, note that $N_i(E_n)$ encodes the Sommerfeld constant ($z_S=4\pi e m_e/h_P^3$)~\cite{Kyritsakis}, where $e$ is the elementary charge, $m_e$ is the electron mass and $h_P$ is the Planck constant.

Without loss of generality, we focus on materials with free-electron-like band structure and no surface states~\cite{Fowler_book}. In this case, for convenience, one can define a new parameter associated with the normal energy measured relative to Fermi level, that is, $\epsilon_n=E_n-\mu$. Therefore, the current density formula, Eq.~(\ref{eq: current_density_exp}), can be written as follows, 
\begin{equation}\label{eq: normal_energy}
J=\int N_i(\epsilon_n)D(\epsilon_n)d\epsilon_n
\end{equation}

It is worth to point out that the effects of finite-temperature on the current density are also included on the $N_i(\epsilon_n)$ term, which is given by a well-grounded formula~\cite{Fowler_book} that can be expressed as,
\begin{equation}\label{eq: temp}
N_i(\epsilon_n)=z_S k_B T \ln \left[1+\exp{\left\{-\epsilon_n/(k_B T)\right\}}\right]
\end{equation}
where $k_B T$ is the Boltzmann factor and $z_S$ is the Sommerfeld constant~\cite{Kyritsakis}, as stated before.

At \textit{zero-temperature} (cold emission) the electrons emitted through field emission are located below the Fermi energy ($\mu$), that is, the only states occupied are those with negative or zero $\epsilon_n$. This situation is  in contrast to the thermionic process for electronic emission, where the electrons must have sufficient energy to escape from the solid~\cite{Neto1}. So, for cold emission, i.e., for $T \rightarrow 0$, the Eq.~(\ref{eq: temp}) is given by,
\begin{equation}\label{eq:10}
N_i(\epsilon_n)_{(T \rightarrow 0K)}=\begin{cases}
-z_S \epsilon_n, & \epsilon<0 \\
0, & \epsilon_n \ge 0 \end{cases}
\end{equation}

Therefore, with the help of Eq.~(\ref{eq:10}), one can rewrite the current density formula, Eq.~(\ref{eq: normal_energy}), as follows,
\begin{equation}\label{eq8}
J=-z_S\int \epsilon_n D(\epsilon_{n}) d\epsilon_{n} ~.
\end{equation}

Note, from Eq.~(\ref{eq8}), that to get the final expression for the current density we still need to determine the electronic transmission coefficient ($D(\epsilon_n)$) before performing the relative integral to the normal component of energy ($\epsilon_n$). In that sense, we will apply the well-established semi-classical (JWKB) approximation~\cite{Murphy, Cutler, Fursey, Neto2, Neto3} to obtain the transmission coefficient. 

\subsection{Barrier penetration coefficient}

As stated before, in order to calculate the PE barrier penetration coefficient ($D(\epsilon_n)$), which can be written in terms of the Gamow exponent ($G$)~\cite{Forbes7}, we use the well-known semi-classical (JWKB) approximation given by~\cite{Murphy, Cutler, Fursey, Neto2, Neto3},
\begin{equation}\label{eq9}
D(\epsilon_{n})= \exp{[-G]}
\end{equation}

It is worth to emphasize that at \textit{zero-temperature}, the NED of the emitted electrons is restricted to the Fermi level over the typical range of electric fields whereupon the field emission occurs. In other words, at \textit{zero-temperature}, the emitted electrons are those that have energy around the Fermi energy, i.e., $E_n \approx \mu$. Therefore, one can expand the exponent of the transmission coefficient ($G$), in Eq.~(\ref{eq9}), in a power series (Taylor-type) and truncate it in second order at the Fermi level in the following manner,
\begin{equation}\label{eq: fermi_level_exp}
G \approx \left.G\right|_F+\left.\left(\frac{\partial G}{\partial \epsilon_n}\right)\right|_F \epsilon_n \equiv G_F+\frac{\epsilon_n}{d_F}
\end{equation}
where the \textit{subscript} $F$ is being used to label parameters evaluated at the Fermi level and the parameter $d_F$ is known as decay width at the Fermi level, as it can be clearly identified by its definition in Eq.~(\ref{eq: fermi_level_exp}).

Replacing Eq.~(\ref{eq: fermi_level_exp}) into Eq.~(\ref{eq9}) yields,
\begin{equation}\label{eq: tunnel_and_density_current}
D(\epsilon_n) \approx D_F \exp{\left(\epsilon_n/d_F\right)}
\end{equation}
where the transmission coefficient $D_F$ for an electron with $\epsilon_n=0$ is given by
\begin{equation}\label{eq: Fermi_level_exp}
D_F\approx\exp[-G_F]=\exp{[-\textit{v}_F b\phi^{3/2}/F]}
\end{equation}

Here, $\textit{v}_F$ is the value, for an electron tunneling with $\epsilon_n=0$, of the barrier form correction factor $\textit{v}$, that in this formulation is entirely general and depends on the barrier of interest~\cite{Forbes6}. In other words, the expression for $\textit{v}_F$ can exhibit different functional dependencies according to the nature of the barrier considered.

Once we get the expression for $D(\epsilon_n)$, Eq.~(\ref{eq: tunnel_and_density_current}), we can integrate out the Eq.~(\ref{eq8}) to obtain the current density.

\subsection{Calculating the field emission current density}

Thus, with the help of Eq.~(\ref{eq: tunnel_and_density_current}), the current density, Eq.~(\ref{eq8}), is given by,
\begin{equation}\label{eq: current_density_replacing}
J=-z_S D_F \int_{-\infty}^{0} \epsilon_n \exp{\left(\epsilon_n/d_F\right)} d\epsilon_n
\end{equation}
where it is known that at \textit{zero-temperature} the upper limit of this integral must be zero since there are no electrons in states above the Fermi level and, for convenience, one can extend the lower limit in the integral to infinity ($\infty$)~\cite{Fowler_book}. 

This procedure simplifies the calculation of the resulting integral in Eq.~(\ref{eq: current_density_replacing}), which leads to the conventional result,
\begin{equation}\label{eq: density_current_conv}
J=J_{conv} \equiv \left(z_S d_F^{2}\right)D_F
\end{equation}

From Eq.~(\ref{eq: density_current_conv}) one can recognize that this result is equivalent to a FN-type equation, and it is a known result that the pre-exponential factor, i.e., the product ($z_S d_F^2$), can be written in a more familiar form,
\begin{equation}\label{eq: alternative}
\left(z_S d_F^{2}\right) \equiv \tau_F^{-2}a \phi^{-1}F^2
\end{equation}
where $a$ is the first FN constant~\cite{Forbes5} and $\tau_F$ is a correction factor (of order unity) defined by Eq.~(\ref{eq: alternative}).

Note that Eq.~(\ref{eq: density_current_conv}) and Eq.~(\ref{eq: alternative}) are general results within the context of smooth planar emitters that do not involve any particular consideration about the form of the tunneling barrier. In the case of the SN barrier used in the \textit{zero-temperature} MG theory, $\tau_F$ is given by the well-known special mathematical parameter $\textit{t}_{MG}$ that appears in MG theory, see Eq.~(\ref{eq5}).

However, in the opposite limit, i.e., for the case that the Fermi energy $\mu$ is comparable with or smaller than the decay width $d_F$, then the mathematical approximation just discussed to solve Eq.~(\ref{eq: current_density_replacing}) is no longer physically valid and the lower limit of integration has to be set at $-\mu$. In this case, integrating out the Eq.~(\ref{eq: current_density_replacing}) leads to a revised FE current density $J_{rev}$ given by,
\begin{equation}\label{eq: Jrev}
J_{rev}\equiv z_S d_F D_F\left[d_F-\mu \exp{\left(-\mu/d_F\right)}-d_F\exp{\left(-\mu/d_F\right)}\right]
\end{equation}

First, note that in the limit $\mu \rightarrow \infty$ we recover the Eq.~(\ref{eq: density_current_conv}), as expected. Moreover, note that one can rewrite the Eq.~(\ref{eq: Jrev}) in a more intuitive form defining a new parameter $\alpha$ as follows,
\begin{equation}\label{eq: alpha}
\alpha \equiv \mu/d_F
\end{equation}

Using Eq.~(\ref{eq: density_current_conv}) and Eq.~(\ref{eq: alpha}) one can rewrite Eq.~(\ref{eq: Jrev}) in the form,
\begin{equation}\label{eq: Jrev2}
J_{rev} \equiv C J_{conv} \equiv \left[1-(1+\alpha)\exp{(-\alpha)}\right] J_{conv}
\end{equation}
where the correction factor $C$ is defined by Eq.~(\ref{eq: Jrev2}).

As before, within the context of smooth planar emitters, this is a general result that does not depend on the explicit form of the barrier considered. 

It is worth to emphasize that the expression Eq.~(\ref{eq: Jrev2}) has two distinct limits. The conventional assumption arises when $\mu>>d_F$, which is equivalent to taking $\alpha >>1$, leading to the limit $C \approx 1$. On the other hand, in the opposite limit, i.e., for $\alpha <<1$, the exponential can be expanded adequately as ($1-\alpha$), which yields the result $C \approx \alpha^2$, and hence,
\begin{equation}\label{eq: j_final}
J_{small \ \mu} \approx \left(\mu/d_F\right)^2 J_{conv}
\end{equation}
or alteratively,
\begin{equation}\label{eq: j_final_2}
J_{small \ \mu} \approx \left(z_S \mu^2\right) D_F
\end{equation}

From Eq.~(\ref{eq: j_final}) or Eq.~(\ref{eq: j_final_2}) we can clearly identify the emergence of a non-FN-type behavior for general planar emitters in the small Fermi energy regime. Furthermore, comparing Eq.~(\ref{eq: j_final_2}) with Eq.~(\ref{eq: density_current_conv}) one can immediately recognize that the main difference between the two distinct regimes for the FE current density is intrinsically related to the pre-exponential factor form. In other words, in the non-FN-type behavior regime the pre-exponential factor form depends on the Fermi energy ($\mu$) only. 

It turns out that the relevance of these results to real materials are directly associated with the typical values of the ratio $\alpha \equiv \mu/d_F$. In that sense, for a metal with work-function $4.5$ eV, $d_F$ typically has values in the range $100$ meV to $250$ meV, whereas for metals Fermi energies are typically around $5$ eV~\cite{Ashcroft-Mermin}, or more. Therefore, we point out that the conventional FN-theory is adequate for metals. This has always been assumed as ``intuitively obvious", but the present work provides a formal mathematical demonstration that this intuition is correct. However, for some non-metals or materials that have small Fermi energy, if FN-type theory is used to describe FE process from such materials, one can observe that the FN-type theory may require that a correction of the type indicated in Eq.~(\ref{eq: j_final}) or Eq.~(\ref{eq: Jrev2}) be taken into account. Note that considerations of this general kind, about Fermi energy values, will also apply if the theory used is more general than conventional FE theory.

\section{Conclusions}
\label{sec:conclusions}

FE process is an essential ingredient for many current technological applications, such as, lamp~\cite{Edgcombe}, X-ray sources~\cite{Edgcombe, Anantram}, FE devices~\cite{Anantram}, and carbon nanotubes~\cite{Zestanakis, Filip, Mikhailovskij}. This phenomena is generally described, as a first approximation, by the standard FN formula, which was derived for planar emitters~\cite{Fowler, Edgcombe} and is expressed in terms of special functions that does not allow a direct physical interpretation, i.e., a numerical approach is required. In that sense, several works have been presented aiming to improve the FN formula as well as take into account effects of tunneling, thermal emission and curvature for field emission in metals~\cite{Cutler, Jensen, Forbes4, Edgcombe, Fischer, Forbes6, Kyritsakis,Deane-Forbes}.

In order to contribute to this issue, in this paper we have analyzed the behavior of the FE current density for a generic tunneling barrier within the context of smooth planar emitters at \textit{zero-temperature}, that is, a planar general tunneling barrier that does not depend on the specific form of the barrier considered.

Without loss of generality, we focused on the materials with free-electron-like band-structure and no surface states. We have applied the semi-classical (JWKB) approximation~\cite{Murphy, Cutler, Fursey, Neto2, Neto3} to obtain the barrier transmission coefficient. We perform a Taylor-type series expansion on the Gamow exponent~\cite{Forbes7} and truncate it in second order at the Fermi level, since at \textit{zero-temperature} the only states occupied are those with energy values below the Fermi level. Then, the general current density formula is obtained by means of integration via the NED.

We present a general evidence of the emergence of a non-FN-type behavior in planar emitters for the case that the Fermi energy ($\mu$) is comparable with or smaller that the decay width ($d_F$), or equivalently, in the limit that the ratio $\alpha = \mu/d_F<<1$. So, for some non-metals or materials that have very small Fermi energy, the FN-type theory may require a revision. However, in the opposite limit, i.e., for $\alpha = \mu/d_F>>1$, we verify that the conventional FN-theory is suitable for usual metals. To sum up, we have shown that there is two distinct behaviors for the FE current density depending on the Fermi energy values.

For completeness, it is worth to emphasize that this kind of non-FN-type behavior has been reported in different scenarios. For example, in multiwall carbon nanotubes~\cite{Zestanakis} as well as in highly curved, nanometer-scale surfaces~\cite{Holgate}. Therefore, it is clear that the relevance of our results for planar emitters depends on the typical values for $\alpha$. For instance, for a usual metal with work function $4.5$ eV, $d_F$ typically has values in the range $100$ meV to $250$ meV, whereas for metals Fermi energies are typically around $5$ eV~\cite{Ashcroft-Mermin}, or more. Thus, one can conclude that the conventional statement is appropriate for metals. However, for some non-metals, the Fermi energy is very much smaller, which implies that the non-FN-type behavior may take place. For such materials, or even for different materials that have small Fermi energy, if FN-type theory is applied, as a first approximation, to describe FE phenomena from such materials, self-consistency may require that the emergence of a non-FN-type behavior presented here might be taken into account. Finally, considerations of this general kind, i.e., related to the Fermi energy values, will also apply if the theory used is more general than conventional FE theory.

\section{Acknowledgments}

The Brazilian agencies \textit{Funda{\c c}\~ao de Amparo \`a Pesquisa do Estado da Bahia} (FAPESB) and \textit{Conselho Nacional de Desenvolvimento Cient\'ifico e Tecnol\'ogico} (CNPq) are acknowledged for partial financial support. The authors would like to thank the professor Dr. Daniel Reyes for carefully reading the manuscript. Moreover, the authors would like to thank the referees for valuable comments on the paper, which certainly contributed to make the paper clearer and more interesting.



\end{document}